\documentclass[pra,aps,twocolumn,superscriptaddress,nofootinbib,nopacs]{revtex4} %---APS---SI---arxiv
\usepackage{amsmath}  \usepackage{amssymb}  \usepackage{amsfonts}  \usepackage{bm}  \usepackage{bbm}   \usepackage{braket}  \usepackage{color}  \usepackage{comment}  \usepackage{dcolumn}  \usepackage{enumerate}  \usepackage{epsfig}  \usepackage{gensymb}  \usepackage{graphicx}  \usepackage{indentfirst}  \usepackage{lmodern}  \usepackage{mathrsfs}  \usepackage{mathtools}  \usepackage{psfrag}  \usepackage{pst-all}  \usepackage{soul}  \usepackage{units}  \usepackage{xcolor}
\usepackage{float} %---APS---SI---arxiv
\usepackage[colorlinks,linkcolor=blue,citecolor=blue,urlcolor=blue,hyperindex,driverfallback=dvipdfm]{hyperref}  \usepackage[T1]{fontenc} %---APS---ACS---SI---arxiv

\def\ii{{\rm i}}  \def\ee{{\rm e}}

        \def\Eb{{\bf E}}  \def\eb{{\bf e}}                            \def\pb{{\bf p}}        \def\rb{{\bf r}}    \def\ub{{\bf u}}   %--- bold vectors
\def\xx{\hat{\bf x}}  \def\yy{\hat{\bf y}}  \def\zz{\hat{\bf z}}            
\begin{document} %---APS---SI---arxiv
\def\bibsection{\section*{\refname}} %---SI---arxiv

%\topmargin 0.0cm  \oddsidemargin 0.2cm  \textwidth 16cm  \textheight 21cm  \footskip 1.0cm \newenvironment{sciabstract}{\begin{quote} \bf} {\end{quote}}  %---Science

% =========================================================
% --- title, affiliations, abstract -----------------------
% =========================================================
\title{Nonlinear Tunable Vibrational Response in Hexagonal Boron Nitride
%\\{\color{gray} \small -- SUPPLEMENTARY INFORMATION -- } %---SI
}

% --- Science affiliations --------------------------------
%\author{Fadil~Iyikanat,$^1$ Andrea~Kone\v{c}n\'{a},$^1$ and F.~Javier~Garc\'{\i}a~de~Abajo$^{1,2\ast}$\\  \\
%\normalsize{$^{1}$ICFO-Institut de Ciencies Fotoniques, The Barcelona Institute of Science and Technology,}\\ \normalsize{08860 Castelldefels (Barcelona), Spain}\\
%\normalsize{$^{2}$ICREA-Instituci\'o Catalana de Recerca i Estudis Avan\c{c}ats, Passeig Llu\'{\i}s Companys 23,}\\
%\normalsize{08010 Barcelona, Spain}\\
%\normalsize{$^\ast$Corresponding author. Email: javier.garciadeabajo@nanophotonics.es}}

% --- OSA affiliations ------------------------------------
%\author[1,2,*]{F.~Javier~Garc\'{\i}a~de~Abajo}
%\affil[1]{ICFO-Institut de Ciencies Fotoniques, The Barcelona Institute of Science and Technology, 08860 Castelldefels (Barcelona), Spain}
%\affil[2]{ICREA-Instituci\'o Catalana de Recerca i Estudis Avan\c{c}ats, Passeig Llu\'{\i}s Companys 23, 08010 Barcelona, Spain}
%\affil[*]{E-mail: javier.garciadeabajo@nanophotonics.es}

% --- APS,SI,arxiv affiliations ---------------------------
\author{Fadil Iyikanat}
\affiliation{ICFO-Institut de Ciencies Fotoniques, The Barcelona Institute of Science and Technology, 08860 Castelldefels (Barcelona), Spain}
\author{Andrea Kone\v{c}n\'{a}}
\affiliation{ICFO-Institut de Ciencies Fotoniques, The Barcelona Institute of Science and Technology, 08860 Castelldefels (Barcelona), Spain}
\author{F.~Javier~Garc\'{\i}a~de~Abajo}
\email{javier.garciadeabajo@nanophotonics.es}
\affiliation{ICFO-Institut de Ciencies Fotoniques, The Barcelona Institute of Science and Technology, 08860 Castelldefels (Barcelona), Spain}
\affiliation{ICREA-Instituci\'o Catalana de Recerca i Estudis Avan\c{c}ats, Passeig Llu\'{\i}s Companys 23, 08010 Barcelona, Spain}

% --- ACS affiliations ------------------------------------
%\author{Fadil Iyikanat}
%\affiliation[ICFO]{ICFO-Institut de Ciencies Fotoniques, The Barcelona Institute of Science and Technology, 08860 Castelldefels (Barcelona), Spain}
%\author{Andrea Kone\v{c}n\'{a}}
%\affiliation[ICFO]{ICFO-Institut de Ciencies Fotoniques, The Barcelona Institute of Science and Technology, 08860 Castelldefels (Barcelona), Spain}
%\author{F.~Javier~Garc\'{\i}a~de~Abajo}
%\email{javier.garciadeabajo@nanophotonics.es}
%\affiliation[ICFO]{ICFO-Institut de Ciencies Fotoniques, The Barcelona Institute of Science and Technology, 08860 Castelldefels (Barcelona), Spain}
%\alsoaffiliation[ICREA]{ICREA-Instituci\'o Catalana de Recerca i Estudis Avan\c{c}ats, Passeig Llu\'{\i}s Companys 23, 08010 Barcelona, Spain}

% --- document format -------------------------------------
%\begin{document} %---ACS
%\ociscodes{XXX} %(300.6530) Ultrafast spectroscopy; (270.1670) Coherent optical effects.} %---OSA
%\date{\today}  \begin{document}  \baselineskip24pt  \maketitle %---Science

% --- abstract --------------------------------------------
\begin{abstract}
%\begin{sciabstract}{\bf
Nonlinear light-matter interactions in structured materials are the source of exciting properties and enable vanguard applications in photonics. However, the magnitude of nonlinear effects is generally small, thus requiring high optical intensities for their manifestation at the nanoscale. Here, we reveal a large nonlinear response of monolayer hexagonal boron nitride (hBN) in the mid-infrared phonon-polariton region, triggered by the strongly anharmonic potential associated with atomic vibrations in this material. We present robust first-principles theory predicting a threshold light field $\sim40\,$MV/m to produce order-unity effects in Kerr nonlinearities and harmonic generation, which are made possible by a combination of the long lifetimes exhibited by optical phonons and the strongly asymmetric landscape of the configuration energy in hBN. We further foresee polariton blockade at the few-quanta level in nanometer-sized structures. In addition, by mixing static and optical fields, the strong nonlinear response of monolayer hBN gives rise to substantial frequency shifts of optical phonon modes, exceeding their spectral width for in-plane DC fields that are attainable using lateral gating technology. We therefore predict a practical scheme for electrical tunability of the vibrational modes with potential interest in mid-infrared optoelectronics. The strong nonlinear response, low damping, and robustness of hBN polaritons set the stage for the development of applications in light modulation, sensing, and metrology, while triggering the search for intense vibrational nonlinear response in other ionic materials.
%}\end{sciabstract}
\end{abstract}

% --- document format -------------------------------------
%\setboolean{displaycopyright}{true} %---OSA
%\begin{document} %---OSA
\maketitle %---APS---OSA---SI---arxiv
\date{\today} %---APS---arxiv
%\tableofcontents %---APS---SI---arxiv optional
%\setcounter{equation}{0} %---OSA
%\setkeys{acs}{maxauthors=0} %---ACS to avoid "et al." in references
%\noindent \textbf{Keywords:} nonlinear response, phonon polaritons, quantum optics, hexagonal boron nitride, mid-infrared spectroscopy %---ACS

% =========================================================
% --- introduction ----------------------------------------
% =========================================================
\section{Introduction} % ... text ... %---APS---OSA---arxiv
%\section*{INTRODUCTION} %---Science

Major improvements in nanofabrication \cite{NLO09,DFB12,DJD20} and colloid chemistry \cite{BCN05,CRJ10} over the last two decades have enabled an extraordinary degree of control over the electromagnetic field associated with light at nanometer scales, well below the free-space light wavelength. These advances have facilitated the development of applications in fields such as optoelectronics \cite{MS16}, optical sensing \cite{NSB15,DSW16,paper308}, photochemistry \cite{C14}, and light harvesting \cite{AP10} by exploiting the linear optical response of the involved materials. Nonlinear effects open up a wealth of possibilities in quantum optics \cite{paper318}, harmonic generation \cite{STV17}, all-optical light modulation and switching \cite{B08_3,G13}, and nonlinear microscopy \cite{CL03,WLN11,DFP12,SMH12_2,HC13} and sensing \cite{NPV09,MMH16,DSW16,DVL17,paper279}. Unfortunately, the nonlinear response of available materials is rather weak, so the manifestation of nonlinear phenomena generally demands their accumulation as light propagates across large structures (i.e., extending over many optical wavelengths \cite{B08_3,G13}), or alternatively, the use of intense optical fields. However, the light intensity that can be tolerated by nanostructures is limited due to heating damage, which can be effectively prevented by reducing the irradiation time, so understandably, progress in nonlinear optics has largely relied on the availability of ultrafast laser pulses with fluences $\lesssim10\,$J/m$^2$.

Two-level atoms constitute the paramount example of a nonlinear system \cite{B08_3} in which resonant photon absorption is prevented after it is formerly excited by a previous photon. However, individual atoms possess small transition strengths, as quantified by the f-sum rule \cite{PN1966}, because only one electron is effectively involved in the excitation, thus resulting in poor coupling to light, and consequently, also a weak nonlinear response. Insulators and semiconductor materials can exhibit larger transition strength associated with a higher density of available valence electrons, and actually, there are notable examples among them offering relatively large nonlinear response combined with low optical absorption that enable applications such as optical parameter oscillators for frequency conversion \cite{B08_3}. The high density of conduction electrons in metals also translates into a larger transition strength, which, although accompanied by substantial optical absorption, still gives rise to relatively strong nonlinearities, particularly in noble metal nanostructures \cite{UKO94,GBG99,RBB07,LTY06,SSN09}.

Localized optical resonances can produce amplification of the externally applied light intensity to enhance nonlinear effects \cite{paper307}. In particular, metal plasmons provide the means to amplify the near field intensity by several orders of magnitude, thus strongly increasing the nonlinear response \cite{LVO05,LTY06,DN07,SO09,BDB10,BBR10,S11_2,HVQ12,KZ12}. Unfortunately, metals suffer from large optical losses \cite{K15_2} that limit their lifetimes, so alternative less lossy materials have been explored, such as graphene, which can sustain electrically tunable \cite{FRA12,paper196} and long-lived \cite{NMS18} plasmons, while simultaneously featuring a strong nonlinear response because of the conic, anharmonic nature of its electronic band structure \cite{M07_2,MZ08}. More precisely, graphene exhibits a large third-order susceptibility that manifests in relatively strong four-wave mixing \cite{HHM10,GPM12,CHC16},
third-harmonic generation (THG)\cite{KKG13,HDP13,SWR18,paper367}, and the optical Kerr effect \cite{ZVB12,DDG16}. Although graphene plasmons have been only observed at low frequencies in the terahertz and mid-infrared domains, they can enhance the intrinsically large nonlinear response of their host material \cite{paper337,paper367}, with potential application in quantum optics \cite{paper318} and high-harmonic generation \cite{paper287}.

The two-dimensionality of graphene is advantageous for nanophotonic devices because it facilitates exposure to the external optical field. Likewise, the vast family of two-dimensional (2D) transition metal dichalcogenides (TMDs) also host interesting nonlinear properties, as revealed by the observation of strong second-harmonic generation (SHG) in odd-layer films of MoS$_{2}$ \cite{MAB13,KNC13}, MoSe$_{2}$ \cite{CCS17},
WS$_{2}$ \cite{JWM14}, and WSe$_{2}$ \cite{WMG15}, as well as THG in MoS$_{2}$ \cite{SKR17,WMP16} and spiral WS$_{2}$ \cite{FJZ17}.

For sufficiently strong external fields, atomic vibrations in solids and molecules can be pushed beyond the harmonic regime, thus exhibiting nonlinear behavior. Surprisingly, harmonic generation associated with atomic vibrations has only been poorly explored, with just a few works focusing on this phenomenon at terahertz frequencies \cite{DYS03,PRG16,NC16,WGS18}, as well as a parallel effort on parametric amplification of optical phonons in SiC \cite{CNF18}. The strength of vibrational nonlinear effects is a key question that determines the range of applications, but this line of research is still open, in search of robust materials with intense response. Mid-infrared nanophotonic devices would benefit from such strong nonlinearity, in particular in 2D platforms that enable easy access to the external field. In this context, hexagonal boron nitride (hBN) emerges as an appealing candidate that exhibits long-lived optical phonon polaritons, although their associated nonlinear response has not yet been assessed.

Here, we find that atomic vibrations in polar crystals can produce strong optical nonlinearities in the mid-infrared spectral region, on par with their electronic counterparts in strongly nonlinear media. We concentrate on monolayer hBN as a material of current interest due to its ability to host long-lived phonon polaritons at spectral bands emerging at around $\sim100\,$meV and $\sim170\,$meV. Through first-principles simulations, we find that the higher-energy band exhibits a substantial degree of asymmetry for atomic vibrations involving stretching of the B-N bond, giving rise to a strongly anharmonic behavior that translates into relatively intense harmonic generation, as well as sizeable Kerr nonlinearity. Phonon polaritons such as those in hBN therefore emerge as a promising platform for mid-infrared nonlinear optics, with applications that include harmonic generation, optical modulation, and quantum blockade at the few-quantum level in nanometer-sized structures, as well as active electrical modulation by applying DC lateral fields.

% --- Figure 1 --------------------------------------------
\begin{figure*}
\centering{\includegraphics[width=1.0\textwidth]{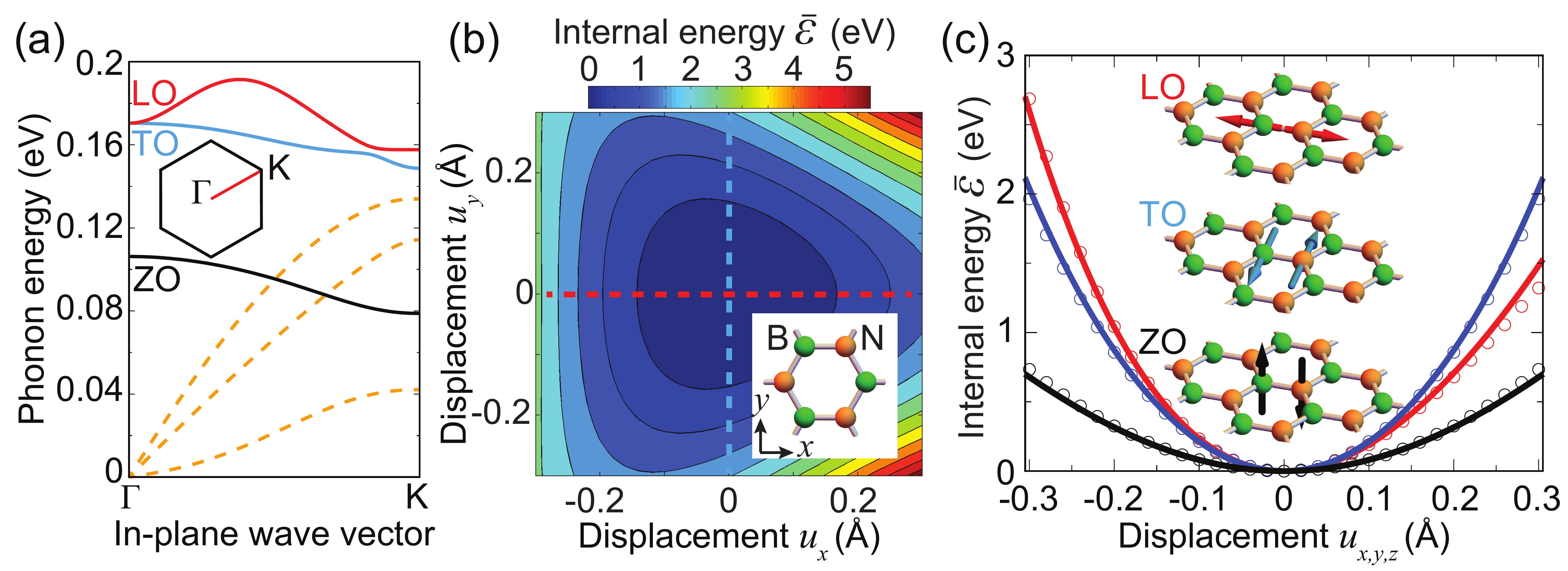}}
\caption{{\bf Anharmonicity in the atomic vibrations of monolayer hBN.} {\bf (a)} Linear dispersion relations of the three acoustic (dashed curves) and three optical (ZO, TO, and LO solid curves) phonon modes along the $\Gamma$-K direction (see first Brillouin zone in the inset) calculated from DFT (see Methods). {\bf (b)} Total internal energy per unit cell as a function of in-plane relative B-N atomic displacement $\ub$ at the $\Gamma$ point. We set the energy origin at the equilibrium configuration $\ub=0$. {\bf (c)} Color-coded cuts along the dashed lines in (b). For comparison, we also show the energy variation associated with out-of-plane motion (ZO mode, black curve and lowest inset). We plot DFT results (symbols) and polynomial fits (curves, see main text). The plot reveals strong non-parabolicity for motion along $x$ (LO mode).}
\label{Fig1}
\end{figure*}

% =========================================================
% --- results ---------------------------------------------
% =========================================================
\section{Results and Discussion}  %... text ... %---APS---OSA---SI---arxiv
%\section*{RESULTS AND DISCUSSION} %---ACS

The dispersion diagram of monolayer hBN phonons contains three relatively high energy optical-phonon bands, one starting at $\sim100\,$meV (ZO) and the other two above $\sim170\,$meV (TO and LO), associated with atomic vibrations primarily involving motion perpendicular and parallel to the hexagonal atomic lattice plane, respectively. In this paper, we study vibrations at the $\Gamma$ point, where the two upper bands are degenerate (Figure\ \ref{Fig1}a). Our treatment is exact when dealing with normally impinging light, but we argue that it also provides a good approximation to model strongly confined phonon-polaritons with in-plane wavelengths down to $\lambda_{\rm p}\sim10$\,nm, whose associated wave vectors $2\pi/\lambda_{\rm p}\sim0.06\,{\AA}^{-1}$ are small compared with the wave vector at the K point $4\pi/3\sqrt{3}a\approx\,1.7{\AA}^{-1}$, where $a=1.446\,{\AA}$ is the B-N bond distance. We thus expect that the linear and nonlinear response functions derived from the present $\Gamma$-point analysis embody an accurate description of the optical properties of mid-infrared phonon-polaritons in monolayer hBN.

At the $\Gamma$ point, atoms in each crystal unit cell follow the same vibration pattern, which can be described in terms of the B-N relative displacement vector $\ub$ according to the equation of motion (see Methods)
\begin{align}
M\left[\ddot{\ub}(t)+\tau^{-1}\dot{\ub}(t)\right]=-\nabla_{\ub}\left[\bar{\mathcal{E}}(\ub)-\;\bar{\pb}(\ub)\cdot\Eb^{\rm ext}(t)\right],
\label{newton3}
\end{align}
where $M=M_{\rm B}M_{\rm N}/(M_{\rm B}+M_{\rm N})$ is the reduced mass, $\bar{\mathcal{E}}(\ub)$ and $\bar{\pb}(\ub)$ are the displacement-dependent configuration energy and dipole per unit cell, respectively, $\Eb^{\rm ext}(t)$ is the electric field of the external light, and we introduce a phenomenological lifetime $\tau$. For concreteness, we take the B-N bond vector along $x$ with the B and N unit-cell atoms placed at $a\,\xx$ and $2a\,\xx$, respectively. For a given $\ub$, the displacements of the two unit cell atoms are $\ub_{\rm B}=-(M/M_{\rm B})\ub$ and $\ub_{\rm N}=(M/M_{\rm N})\ub$ (i.e., positive $u_x$ corresponds to stretching), where $M_{\rm B}=10.811\,$Da and $M_{\rm N}=14.007\,$Da are the average masses corresponding to the natural isotopic abundances of these two elements. In what follows, we set $\tau=2\,$ps, which is consistent with the lifetimes observed in optical measurements \cite{GDV18}. In addition, we calculate $\bar{\mathcal{E}}(\ub)$ and $\bar{\pb}(\ub)$ using density-functional theory (DFT), as explained in the Methods section. Because we focus on the $\Gamma$ point, atomic displacements preserve translational crystal symmetry, so DFT methods for infinite crystals can be straightforwardly applied. We concentrate on the upper optical phonon branches, associated with in-plane atomic motion (i.e., $u_z=0$). Retaining only up to quartic terms in $\bar{\mathcal{E}}$ and cubic terms in $\bar\pb$ compatible with mirror symmetry relative to the $u_y=0$ line and three-fold crystal symmetry around $\zz$, our DFT calculations for $u\le0.03\,{\AA}$ lead to the fitted expressions
\begin{subequations}
\label{Epu}
\begin{align}
\bar{\mathcal{E}}(\ub)&\approx e_0\,(u_x^2+u_y^2)+e_1\,u_x(u_x^2-3u_y^2)+e_2\,u^4, \label{Uu} \\
\bar{\pb}(\ub)&\approx \bar{p}_0\,\xx+Q_0\,\ub+Q_1\big[(u_x^2-u_y^2)\,\xx-2\,u_xu_y\,\yy\big]+Q_2\,u^2\,\ub, \label{pu}
\end{align}
\end{subequations}
with coefficients $e_0=0.2176$, $e_1=-0.1126$, $e_2=0.0489$, $\bar{p}_0=-0.6756$, $Q_0=-0.346$, $Q_1=-0.069$, and $Q_2=0.110$ expressed in atomic units. We note that $\pb(\ub)$ is accompanied by the external field in eq\ \ref{newton3}, so both of the expansions in eqs\ \ref{Epu} account for corrections up third order in the external field. As a result of the crystal symmetries noted above, the energy landscape (Figure\ \ref{Fig1}b) exhibits a more anharmonic profile for motion along $x$, as clearly observed when comparing cuts across $u_x=0$ and $u_y=0$ (Figure\ \ref{Fig1}c). For completeness, we calculate (with coefficients in atomic units) $\bar{\mathcal{E}}\approx0.0839\,u_z^2-0.0167\,u_z^4$ and $\bar{p}_z\approx-0.304\,u_z+0.194\,u_z^3$ for out-of-plane motion at $u_x=u_y=0$; these expressions reveal a more harmonic potential (i.e., smaller nonlinear effects) and a light-coupling dipole of similar strength in the ZO mode.

Equations\ \ref{Epu} encapsulate all the information that is needed to study the vibrational dynamics driven by external illumination in the spectral region near the upper optical modes at the $\Gamma$ point according to eq\ \ref{newton3}. In particular, the corresponding unperturbed in-plane mode energy $\hbar\omega_0=\hbar\sqrt{2e_0/M}\approx170\,$meV is in excellent agreement with previous theoretical \cite{SGC17} and experimental \cite{RHS97,CSF17} results. Incidentally, the permanent dipole $\bar{p}_0$ does not affect optical phonons at the $\Gamma$ point, although it contributes to the dynamics of acoustic modes.

In what follows, we study the response to a monochromatic external field $\Eb^{\rm ext}(t)=2\Eb_0\cos(\omega t)$ of frequency $\omega$ by solving eq\ \ref{newton3} either perturbatively or in the time domain, thus yielding the time-dependent displacement vector $\ub(t)$, and from here the unit-cell induced dipole $\bar\pb(\ub)-\bar\pb(0)$, from which we extract the nonlinear response functions of monolayer hBN.

% --- Figure 2 --------------------------------------------
\begin{figure*}
\centering{\includegraphics[width=1.0\textwidth]{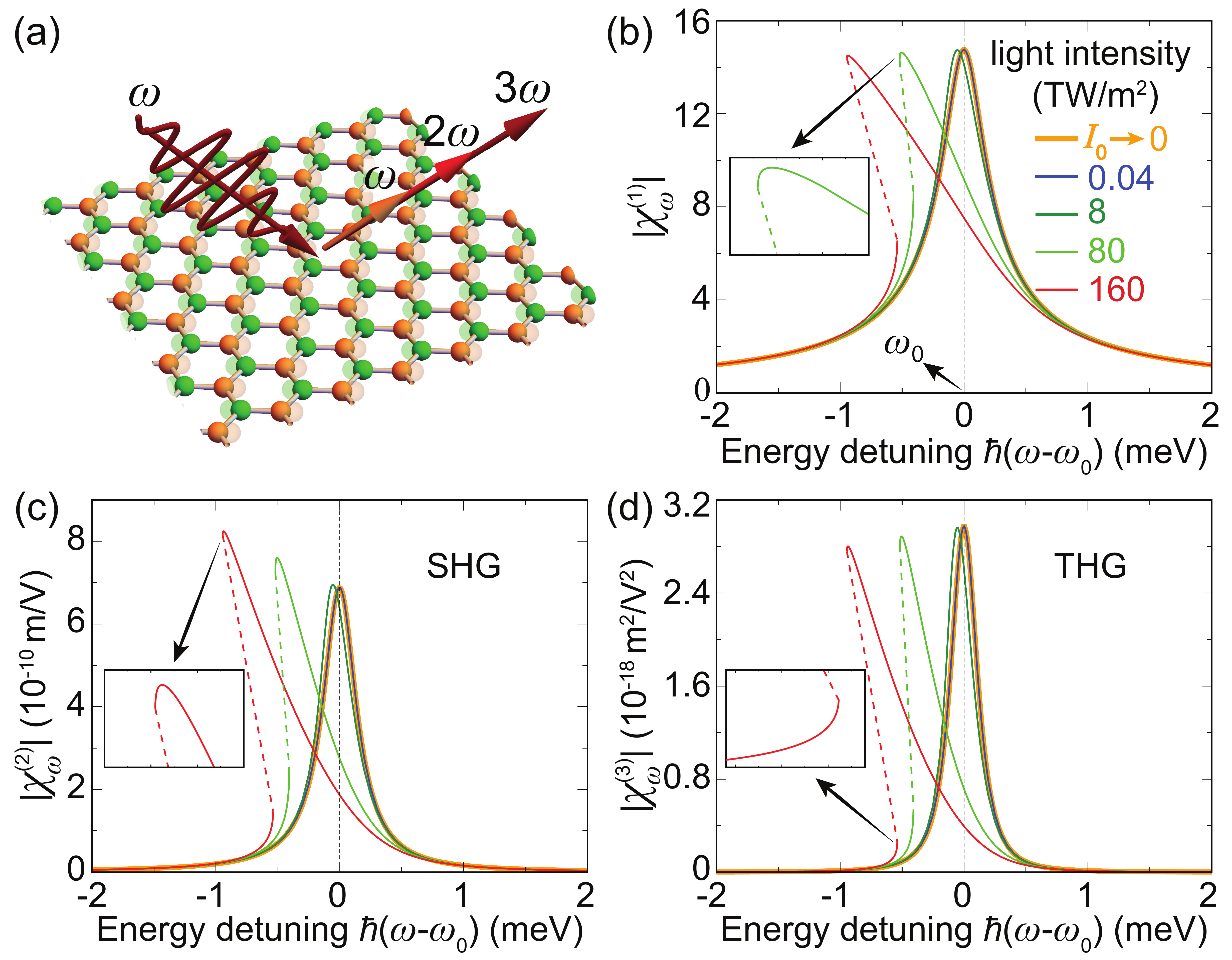}}
\caption{{\bf Nonlinear optical response of monolayer hBN in the upper optical-phonon region.} {\bf (a)} We consider normally impinging light with linear polarization along one of the B-N bond directions $x$. {\bf (b-d)} Spectral dependence of the susceptibilities $\chi^{(s)}_\omega$ associated with the fundamental (b), SHG (c), and THG (d) frequencies ($s=1-3$, respectively) for a series of increasing incident light intensities (see legend in (b)). We show the low-intensity perturbative limit $\chi^{ss}_\omega$ for comparison (thick orange curves). The photon energy $\hbar\omega$ is referred to the linear mode resonance ($\hbar\omega_0\approx170\,$meV). We assume a phonon lifetime of $\tau=2\,$ps ({\it i.e.}, $\hbar\tau^{-1}\approx0.33\,$meV).}
\label{Fig2}
\end{figure*}

\subsection*{Nonperturbative Nonlinear Response in Monolayer hBN}

Based on the energy landscapes shown in Figure\ \ref{Fig1}b,c, we expect a strong nonlinear response associated with atomic vibrations for in-plane displacement vectors oriented along $x$ (the unit-cell B-N bond direction). Because this is a symmetry axis, such vibrations are rigorously constrained to $u_y=u_z=0$ if the external field is also oriented along $x$, so inserting eqs\ \ref{Epu} into eq\ \ref{newton3} and plugging external monochromatic light of frequency $\omega$, the equation of motion reduces to
\begin{align}
&\ddot{u}_x+\tau^{-1}\dot{u}_x+\omega_0^2u_x+(3e_1/M)\,u_x^2+(4e_2/M)\,u_x^3 \nonumber\\
&=(2/M)(Q_0+2Q_1u_x+3Q_2u_x^2)\,E_0\cos(\omega t), \label{newtonx}
\end{align}
which is a generalization of the Duffing equation \cite{TYP17}. In Figure\ \ref{Fig2}, we present results obtained by numerically integrating this equation as a function of $\omega$ for different levels of light intensity $I_0=2E_0^2/Z_0$, where $Z_0=\sqrt{\mu_0/\epsilon_0}\approx376.73\,\Omega$ is the vacuum impedance. From the solution for the time-dependent displacement $u_x(t)$, we express the induced dipole (eq\ \ref{pu}) as $p_x(t)=Q_0u_x(t)+Q_1u_x^2(t)+Q_2u_x^3(t)$. In practice, we solve the above differential equation starting from some initial boundary conditions (see below) and integrating up to a large time $t=N\tau_\omega\gg\tau$ (expressed as a multiple of the optical period $\tau_\omega=2\pi/\omega$), so that the transient response produced after plugging the external light is attenuated to a negligible level. We then compute the susceptibility associated with a harmonic $s$ as
\begin{align}
\chi^{(s)}_\omega=\frac{1}{\epsilon_0(E_0)^s\mathcal{V}\tau_\omega}\int_{N\tau_\omega}^{(N+1)\tau_\omega}dt\;p_x(t)\;\ee^{\ii s\omega t},
\label{chisw}
\end{align}
where $\epsilon_0$ is the vacuum permittivity and we divide by a volume $\mathcal{V}=\mathcal{A} h$ given by the product of the unit cell area $\mathcal{A}=(3\sqrt{3}/2)\,a^2=5.43\,{\AA}^2$ and the layer thickness $h=3.3\,{\AA}$, with the latter approximated to the interatomic plane distance in bulk hBN.

At low intensities, the susceptibilities exhibit a Lorentzian profile of decreasing width as the harmonic order $s$ increases. Our numerical results converge well to the perturbative analytical limit (see Methods) for $I_0\rightarrow0$ (thick orange curves in Figure\ \ref{Fig2}b-d). When the intensity increases, the spectral peak is red shifted, and eventually, we reach a region of bistability. We explore this behavior by using the converged solution for each $\omega$ as the initial condition to calculate the response for a slightly different $\omega$; this leads to two branches, corresponding to increasing or decreasing frequency starting from -2\,meV or 2\,meV detuning, respectively. Such behavior is observed for all harmonics investigated in Figure\ \ref{Fig2}. In the bistability region, a third unstable branch exists \cite{B08_3}, which we illustrate by dashed lines, introduced here as a guide to the eye. Nonpertubative effects are perceptible for light intensities $I_0\gtrsim10\,$TW/m$^2$ (i.e., $E_0\gtrsim40\,$MV/m).

% --- Figure 3 --------------------------------------------
\begin{figure*}
\centering{\includegraphics[width=1.0\textwidth]{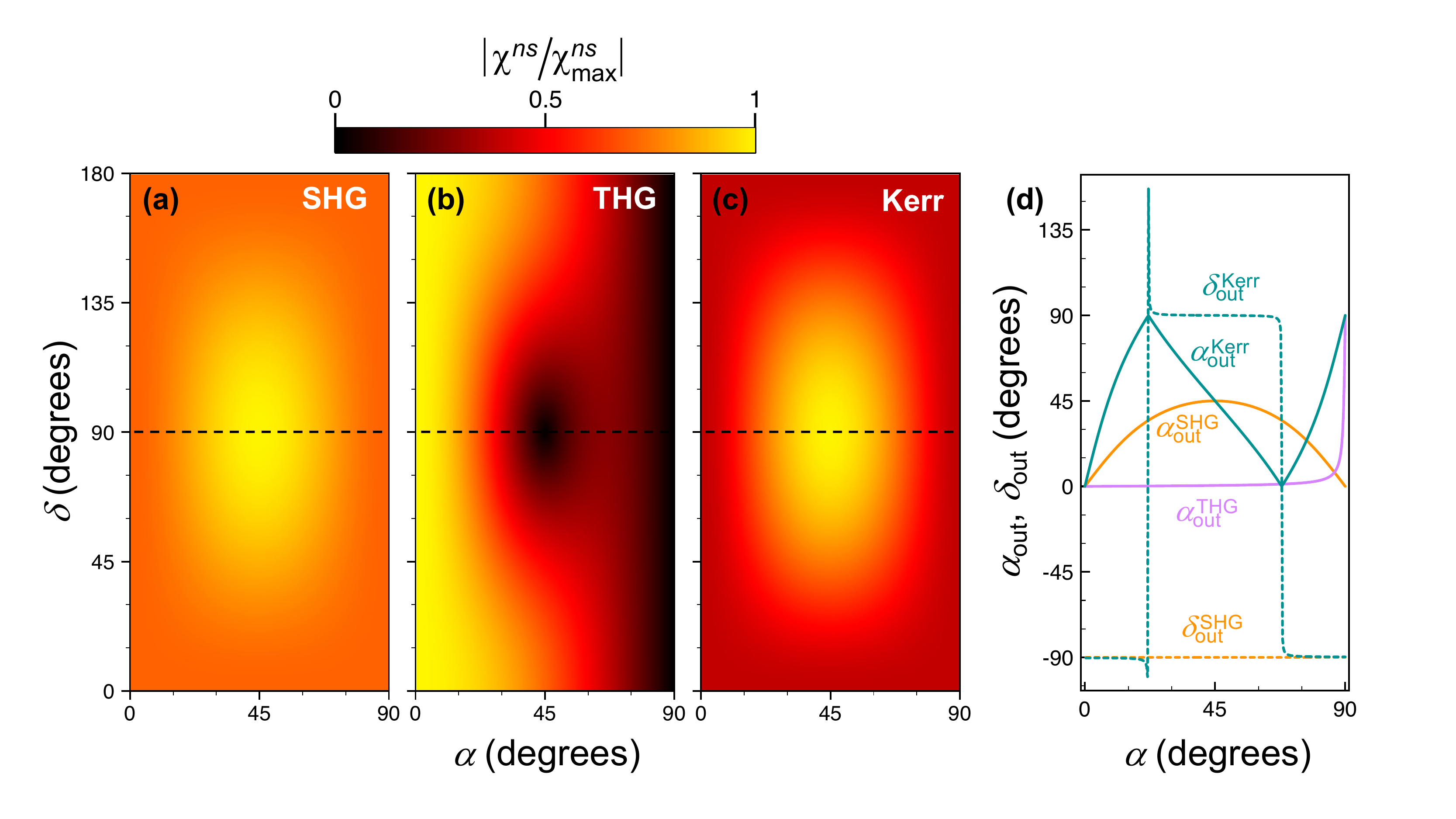}}
\caption{{\bf Polarization dependence of the nonlinear response.} {\bf (a-c)} Perturbative results for the dependence of the nonlinear susceptibility associated with SHG, THG, and the Kerr effect on the polarization of the external light under normal incidence on monolayer hBN. The polarization angles define the field amplitude as $\cos\alpha\,\xx+\sin\alpha\,\ee^{\ii\delta}\yy$. Maxima in the color scale correspond to $\chi^{22}_{\rm max}=9.7\times10^{-10}\,$m/V, $\chi^{33}_{\rm max}=3.0\times10^{-18}\,$m$^2$/V$^2$, and $\chi^{31}_{\rm max}=8.3\times10^{-15}\,$m$^2$/V$^2$ for the SHG (a), THG (b), and Kerr (c) susceptibilities, assuming a polariton lifetime $\tau=2\,$ps. {\bf (d)} Polarization angles of the output nonlinear components as a function of $\alpha$ for $\delta=90^\circ$ (i.e., along the dashed lines in (a-c)).}
\label{Fig3}
\end{figure*}

\subsection*{Polarization Dependence}

\begin{figure*}
\centering{\includegraphics[width=1.0\textwidth]{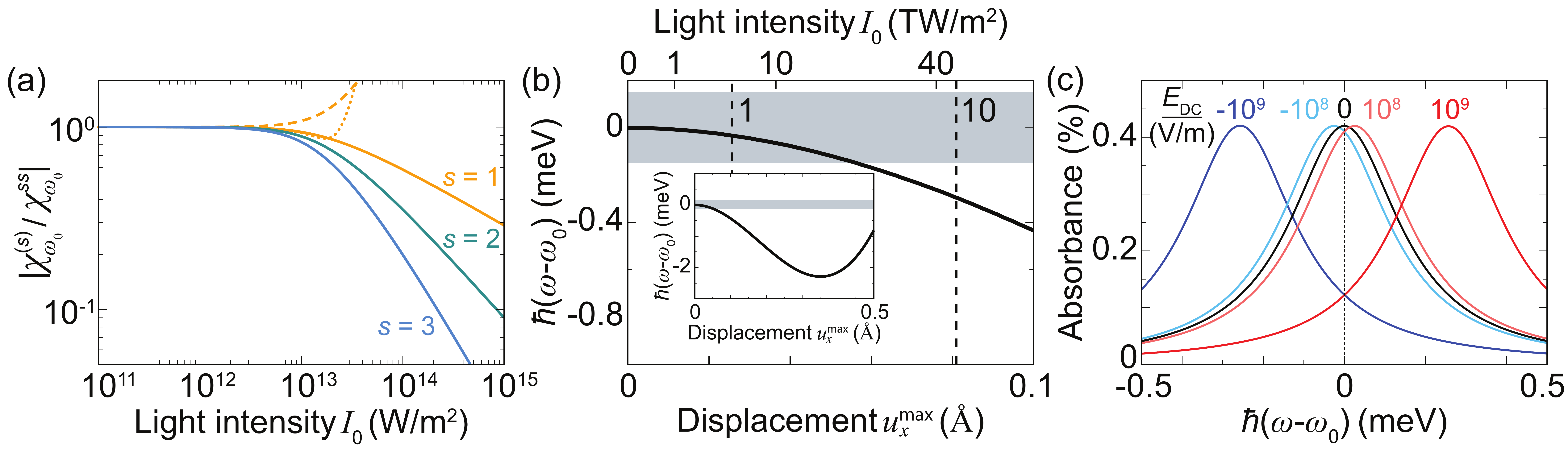}}
\caption{{\bf Nonlinear saturation of the optical response.} {\bf (a)} On-resonance susceptibilities $\chi^{(s)}_{\omega_0}$ associated with the fundamental ($s=1$), SHG ($s=2$), and THG ($s=3$) response as a function light intensity $I_0$ under the conditions of Figure\ \ref{Fig2}, normalized to the low-$I_0$ limit $\chi^{ss}_{\omega_0}$. Analytical perturbative results are shown as broken curves, including the contribution of $\chi_{\omega_0}^{31}$ (dashed curve) and $\chi_{\omega_0}^{51}$ (dotted curve). {\bf (b)} Amplitude dependence of the in-plane vibration frequency for motion along $x$. The lower horizontal axis corresponds to the maximum displacement, while the upper axis indicates the light intensity needed to reach it according to Figure\ \ref{Fig2}b. The two dashed vertical lines indicate the $\sqrt{2}$ times the rms displacement in a $\sim1\,$nm$^2$ island (18 unit cells) for a phonon population of 1 and 10 phonons, as indicated by labels. The shaded region indicates a FWHM corresponding to a lifetime $\tau=2\,$ps. {\bf (c)} Spectral change in the linear absorbance spectrum of monolayer hBN produced by a uniform in-plane DC field along $x$ for different values of the field amplitudes.}
\label{Fig4}
\end{figure*}

The in-plane anisotropy of hBN translates into a strong dependence of the nonlinear response on light polarization, which we analyse in Figure\ \ref{Fig3}. Specifically, we represent the perturbative susceptibilities associated with SHG, THG, and the Kerr effect (see Methods) for normally impinging light of frequency $\omega=\omega_0$ tuned to the in-plane $\Gamma$ phonon frequency as a function of polarization angles $(\alpha,\delta)$, defined in such a way that the incident field amplitude vector is proportional to $\cos\alpha\,\xx+\sin\alpha\,\ee^{\ii\delta}\yy$. The second-harmonic response (Figure\ \ref{Fig3}a) is independent of the direction of the field amplitude for linear polarization ($\delta=0$), while an absolute maximum is observed for circularly polarized light (CPL, $\alpha=45^\circ$, $\delta=90^\circ$) with a relative enhancement of 41\%. Interestingly, the polarization of the SHG signal under CPL irradiation is reversed (output polarization angles $\alpha_{\rm out}=45^\circ$, $\delta_{\rm out}=-90^\circ$; see Figure\ \ref{Fig3}d). We find that THG (Figure\ \ref{Fig3}b) is maximum for polarization along $x$ ($\delta=0$), where it is completely depleted for CPL, in agreement with the intuitive conclusions extracted from the anharmonicity observed in Figure\ \ref{Fig1}c for oscillations parallel or perpendicular with respect to the B-N bond direction. We also analyze the third-order Kerr susceptibility (Figure\ \ref{Fig3}c), which is maximal for CPL, in which case the polarization angles of the nonlinear Kerr response are the same as the incident ones (Figure\ \ref{Fig3}d). Incidentally, right on resonance ($\omega=\omega_0$), we find that $\chi^{31}_{\omega_0}$ is $90^\circ$ out of phase relative to both $\chi^{11}_{\omega_0}$ and $\chi^{51}_{\omega_0}$ (see Methods), so the relative correction to the polarization intensity coming from the Kerr effect scales with the incident intensity as $I_0^2$, with contributions at that order arising from both ${\rm Re}\{\chi^{11*}_{\omega_0}\chi^{51}_{\omega_0}\}$ ({\it i.e.}, through mixing with the linear amplitude) and $|\chi^{31}_{\omega_0}|^2$. This translates into a dependence of $|\chi^{(1)}|^2$ on $I_0$ as shown in Figure\ \ref{Fig4}a (see also discussion in Methods), where the lowest-order correction (dashed curve for $s=1$, obtained by including $\chi^{31}_{\omega_0}$) leads to the wrong sign in the variation of the first-harmonic intensity, whereas the addition of $\chi^{51}$ (dotted curve) produces an initial depletion, in agreement with the nonperturbative result (solid curve), although this approximation eventually breaks down for larger incident intensity.

% --- Figure 4 --------------------------------------------

\subsection*{Saturation and Quantum Blockade}

The onset of saturation at $\sim10\,$TW/m$^2$ is illustrated in Figure\ \ref{Fig4}a, where we plot the susceptibilities associated with polarization emerging at the fundamental, SHG, and THG frequencies for on-resonance illumination at $\omega=\omega_0$. Saturation occurs faster as the harmonic order increases because this involves higher powers of the fundamental field amplitude. From a physical viewpoint, this plot essentially describes the combination of anharmonic oscillations and the effective coupling strength to external light. We find the first of these factors to be of interest {\it per se} because it affects the departure from harmonic behavior at the few-quanta level. In fact, this is the basis for the quantum blockade phenomenon, which we mentioned above for two-level systems: for a sufficiently anharmonic response, the energy of a two-quanta state differs from twice the energy of one quantum. Quantum blockade has been observed in cavity quantum electrodynamics experiments, whereby a two-level atom is coupled to an optical cavity \cite{BBM05}, so that the system inherits a strong anharmoniticy from the former, as well as a large coupling to light from the latter. Incidentally, this type of effect has also been theoretically studied with graphene-plasmon cavities \cite{paper184}, a configuration in which Rabi vacuum splitting can be discernible \cite{paper176}, with a view to realizing quantum-optics devices in a solid-state environment by benefiting from the strong nonlinearity of this material. However, the fabrication of few-nanometer-sized graphene structures capable of sustaining high-quality plasmons remains an experimental challenge.

Atomic vibrations in monolayer hBN provide an excellent alternative to realize quantum blockade in compact structures, which can profit from the structural stability of this material, as well as from the long lifetime and optical strength of its phonon polaritons. We explore this possibility by estimating the oscillation frequency associated with a given maximum displacement (Figure\ \ref{Fig4}b) (see Methods). Larger oscillation amplitudes initially lead to frequency redshifts as a result of the reduction in the interatomic potential relative to a perfect parabola, quantified through the $e_1<0$ term in eq\ \ref{Uu}. We also indicate in this figure an estimate for the root mean square (rms) amplitude associated with the the in-plane optical phonon mode in a $\sim1\,$nm$^2$ hBN island (18 unit cells) for an occupation of either 1 or 10 phonons (see Methods; the rms amplitude is proportional to $\sqrt{n/A}$, where $n$ is the phonon occupation number and $A$ is the area of the island). Incidentally, we multiply the rms displacement by $\sqrt{2}$ to compare with the maximum displacement used in the horizontal axis of Figure\ \ref{Fig4}b. The latter produces a frequency shift that exceeds the FWHM of the resonance assuming a lifetime $\tau=2\,$ps, therefore indicating the onset of quantum blockade.

\subsection*{Electrical Tunability}

A lateral DC field acting on the hBN monolayer along $x$ produces a change in the B-N bond distance to minimize energy. We argue that the strength of the in-plane DC field that can be applied through lateral gating can reach $\sim10^9\,$V/m, which is one order of magnitude larger than the maximum optical field considered in Figure\ \ref{Fig2}. Still, a resonant optical field of amplitude $2E_0\cos(\omega_0t)$ induces a maximum atomic displacement $\approx(2Q_0\tau/M\omega_0)E_0$ assisted by the amplifying mechanism of spring motion, while the displacement due to a DC field of the same magnitude is a factor of $\omega_0\tau\sim500$ smaller. Nevertheless, we show next that the effect is strong enough to shift the phonon resonance by more than its spectral width, therefore enabling a practical route towards electrical light modulation that could find application in optoelectronics. We start our analysis from eq\ \ref{newtonx} by substituting the applied field by $E_{\rm DC}+2E_0\cos(\omega_0t)$. The DC component $E_{\rm DC}$ can be readily absorbed in a new set of parameters $\omega_0$, $e_1$, $Q_0$, and $Q_1$, from which only the variation of $\omega_0$ produces a sizeable effect. Obviously, no constant force term can remain in eq\ \ref{newtonx}, a condition from which we find an equilibrium displacement $u_x^0\approx Q_0E_{\rm DC}/M\omega_0^2$. Because of the lack of parabolicity of the confining potential (see Figure\ \ref{Fig2}c), we expect a shift in the resonance frequency of in-plane phonon at the $\Gamma$ point; after some algebra, we find $\Delta\omega_0\approx(3Q_0e_1/M\omega_0^2-2Q_1)E_{\rm DC}/M\omega_0$, which is linear in the DC field and reaches $\sim0.26\,$meV for $E_{\rm DC}\sim10^9\,$V/m. We also find that  $E_{\rm DC}^2$ corrections amount to less than 1\%, whereas the linear and nonlinear optical responses of the material just experience a rigid frequency shift by $\Delta\omega_0$, with their magnitudes remaining nearly unaffected. This is illustrated by examining the absorbance of monolayer hBN ($\approx(4\pi\epsilon_0h\omega/c){\rm Im}\{\chi^{(1)}_\omega\}$), which reveals a peak shift by nearly twice the spectral width when the DC field is varied in the $\pm10^9\,$V/m range (Figure\ \ref{Fig4}c). Therefore, we anticipate that lateral gating can be used as an efficient mechanism for light modulation in the mid-infrared regime using hBN vibrational modes.

% --- Figure 5 --------------------------------------------
\begin{figure*}
\centering{\includegraphics[width=1.0\textwidth]{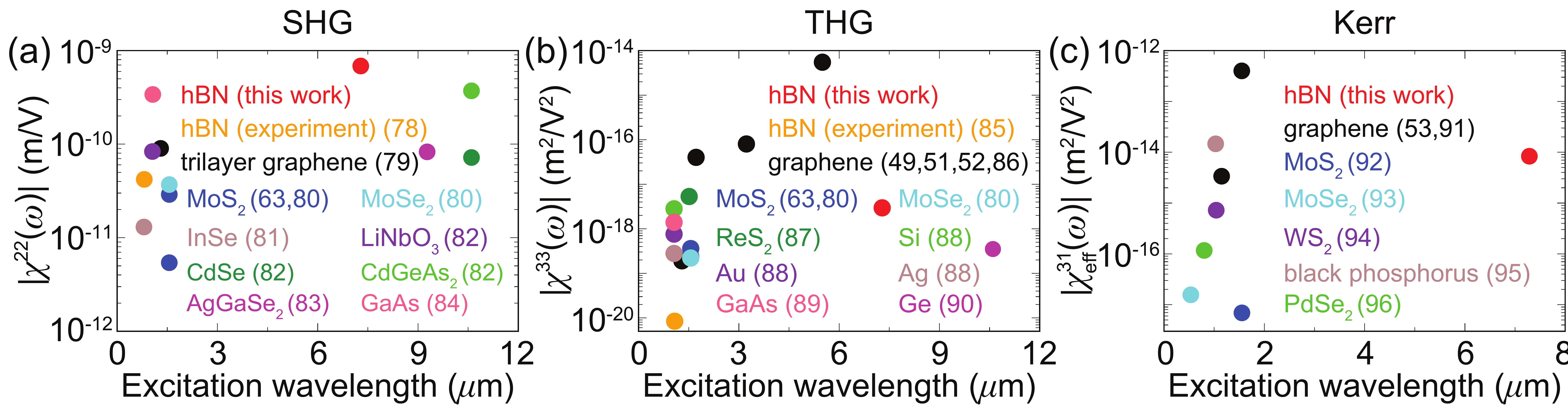}}
\caption{{\bf The nonlinear optical response of hBN in context.} We compare the SHG, THG, and Kerr effect susceptibilities measured in different materials \cite{KFG19,SLH18,paper316,WMP16,HYS19,N05,HK97,SKK97,PAF21,paper367,KKG13,SWR18,JHC18,CMS17,BBM1969,BB1971,WPT1980,ZVB12,DBB16,LXW15,WFC14,DLZ16,YAL18,JWY20} (see color-matching reference numbers in the legends) with our calculations for monolayer hBN at the in-plane $\Gamma$ phonon frequency.
%{\red (1) Kim \textit{et al.}\cite{KFG19}, (2) Shan \textit{et al.}\cite{SLH18} (3) Autere \textit{et al.}\cite{paper316}, (4) Woodward \textit{et al.}\cite{WMP16}, (5) Hao \textit{et al.}\cite{HYS19}, (6) Nikogosyan \textit{et al.}\cite{N05}, (7) Harasaki \textit{et al.}\cite{HK97}, (8) Shoji \textit{et al.}\cite{SKK97}, (9) Popkova \textit{et al.}\cite{PAF21}, (10) Calafell \textit{et al.}\cite{paper367}, (11) Kumar \textit{et al.}\cite{KKG13}, (12) Soavi \textit{et al.}\cite{SWR18}, (13) Jiang \textit{et al.}\cite{JHC18}, (14) Cui \textit{et al.}\cite{CMS17}, (15) Bloembergen \textit{et al.}\cite{BBM1969}, (16) Burns \textit{et al.}\cite{BB1971},(17) Watkins \textit{et al.}\cite{WPT1980}, (18) Zhang \textit{et al.}\cite{ZVB12}, (19) Demetriou \textit{et al.}\cite{DBB16}, (20) Liu \textit{et al.}\cite{LXW15}, (21) Wang \textit{et al.}\cite{WFC14}, (22) Dong \textit{et al.}\cite{DLZ16}, (23) Yang \textit{et al.}\cite{YAL18}, and (24) Jia \textit{et al.}\cite{JWY20}.}
}
\label{Fig5}
\end{figure*}

% =========================================================
% --- conclusion ------------------------------------------
% =========================================================
\section{Concluding Remarks}  %... text ... %---APS---OSA---SI---arxiv
%\section*{CONCLUDING REMARKS} %---ACS

In conclusion, we reveal monolayer hBN as an excellent nonlinear material at frequencies determined by its optical phonons. We base our results on first-principles predictive theory for the potential energy surface and induced dipole density in the material as a function of atomic positions. The optical response associated with atomic vibrations in monolayer hBN contains a substantial anharmonic component that gives rise to relatively intense second- and third-harmonic generation, as well as Kerr nonlinearity, as we show in comparison to existing experimental results for other 2D materials (Figure\ \ref{Fig5}). We stress the fact that, in contrast to the nonlinear response arising from the electronic degrees of freedom, atomic vibrations offer a more robust platform with a lower level of optical losses. In particular, hBN can be compared with graphene, which operates in the same spectral range, but suffers from intrinsic losses that limit the external light intensities that can be applied without producing material damage. Atomic vibrations in hBN are indeed immune to strong optical absorption, such as that taking place in metallic systems, which leads to an elevation of the electronic temperature and an associated change in the optical response (i.e., an incoherent form of nonlinearity) that can mask and reduce the strength of coherent effects. Although graphene shows a larger nonlinear response associated with electronic degrees of freedom (Figure\ \ref{Fig5}), we find hBN to be second best, and we argue that the vibrational origin of its optical response should permit elevating the applied light intensity with less heating of the material. We have focused on vibrations around the in-plane optical mode frequency in hBN, but we anticipate that future studies will explore other materials, covering a wide range of mid-infrared frequencies, and possibly hosting strongly nonlinear vibrational resonances.

Edges in finite hBN islands and defects in actual samples can modify the phonon characteristics, for example by producing localization at atomic scales \cite{HRK20}, which could affect the nonlinear response. Another interesting avenue consists in introducing strong in-plane DC fields to actively modify the nonlinear response, which we have predicted to enable phonon shifts exceeding their spectral width. In this respect, the insulator character of hBN should enable the presence of large lateral DC fields greatly exceeding those that are attainable through optical illumination. We have left aside thermal effects, which could also modify the nonlinear response, in particular in view of the fact that the in-plane mode population is $\sim1$ at the melting temperature of hBN ($>2900$\,K). The strong anharmonic response of hBN could be enhanced through resonant nanostructures \cite{paper307}, a possibility that deserves further exploration to assess the prospects for mid-infrared nonlinear nanophotonics, which could benefit from the strong interest that this material is currently attracting in the scientific community.

The isotopic purity of the material influences the phonon lifetime, and thus requires further investigation in connection to nonlinear effects. We have made some emphasis on the separation between the intrinsic anharmonic motion (i.e., the deviation in the potential energy surface from a parabolic profile) and the optical strength of the optical phonons (i.e., the dipole moment associated with atomic displacements). While the overall nonlinear response is the combined result of both of these aspects, we argue that the former needs to be examined separately, as it controls the possibility of having quantum blockade, whereby subsequent excitation is prevented by the nonlinear effects produced in response to previous excitations. In monolayer hBN, we find that nonlinear response at the few-quanta level is feasible by using structures of 1\,nm lateral size, thus holding the potential for realizing quantum gates based on mid-infrared atomic vibrations in a robust solid-state material platform.

% =========================================================
% --- methods/appendix ------------------------------------
% =========================================================
%\section*{METHODS} %---ACS optional
\begin{widetext} %---arxiv
\appendix %---APS---OSA---SI---arxiv optional
\renewcommand{\thesection}{A} %---SI---arxiv optional
\renewcommand{\theequation}{A\arabic{equation}} %---SI---arxiv optional
%\setcounter{equation}{0} %---OSA optional

% --- text for methods/appendix ----------------------------
%\noindent{\bf Title Format.} ... text ... %---ACS format for Methods section headers
%\section{title} ... text ... %---APS---SI---arxiv format for Appendix section headers (not for OSA)
%\subsection{title} ... text ... %---APS---OSA---SI---arxiv format for Appendix subsection headers
%\renewcommand{\theequation}{A\arabic{equation}} %---OSA for A, B, ... subsections of Appendix

% --- nonlinear theory ------------------------------------
\section{Nonlinear Vibrational Optical Response} We examine the linear and nonlinear optical response associated with the atomic vibrations of monolayer hBN. The atomic scale under consideration is small compared with the light wavelength, so we work in the electrostatic limit and introduce the external light through an optical scalar potential $\phi^{\rm ext}(\rb,t)$ acting on the B and N atoms (labeled by an index $l$), which oscillate around their equilibrium positions with time-dependent displacements $\ub_l(t)$. For each configuration, defined by the set of the atomic displacements $\{\ub\}$, we use DFT to calculate both the internal energy $\mathcal{E}(\{\ub\})$ and the charge density distribution $\rho(\{\ub\},\rb)$ (see below). Adopting the Born-Oppenheimer approximation to separate electronic and vibrational motions, and describing the latter classically, we write the Lagrangian of the system as $\mathcal{L}=(1/2)\sum_l M_l|\dot{\ub}_l(t)|^2-\mathcal{E}(\{\ub\})-\int d^3\rb\;\rho(\{\ub\},\rb)\;\phi^{\rm ext}(\rb,t)$, where $M_l$ denotes the mass of atom $l$, whereas the integral term stands for the potential energy due to the interaction with the external potential. From the Lagrange equation $\partial_t\nabla_{\dot{\ub}_l}\mathcal{L}=\nabla_{\ub_l}\mathcal{L}$, we find
\begin{align}
M_l\left[\ddot{\ub}_l(t)+\tau^{-1}\dot{\ub}_l(t)\right]=-\nabla_{\ub_l}\left[\mathcal{E}(\{\ub\})+\int d^3\rb\;\rho(\{\ub\},\rb)\;\phi^{\rm ext}(\rb,t)\right],
\label{newton1}
\end{align}
where we have introduced a phenomenological lifetime $\tau$ ($=2\,$ps in our calculations).

In the linear regime, we can approximate $\nabla_{\ub_l}\mathcal{E}(\{\ub\})\approx\sum_{l'}\mathcal{D}_{ll'}\cdot\ub_{l'}$ around the equilibrium configuration $\{\ub=0\}$, where $\mathcal{D}_{ll'}$ is the so-called dynamical matrix. The eigenvalues of this matrix define the phonon dispersion relations, as presented in Figure\ \ref{Fig1}a for monolayer hBN based on our DFT calculations for $\mathcal{D}_{ll'}$ (see below).

The wave vectors associated with far-field light or even tip-based illumination ($\sim1/R_{\rm tip}<0.1\,$nm$^{-1}$ for a typical tip radius $R_{\rm tip}>10$\,nm) are at least two orders of magnitude smaller than the reciprocal lattice vectors ($\ge4\pi/3a\approx29\,$nm$^{-1}$, where $a=1.446$\,{\AA} is the B-N bond distance), so for applications in photonics, we are generally interested in atomic vibrations close to the $\Gamma$ point (see Figure\ \ref{Fig1}a). We thus study vibrations at this point as a good approximation to understand the nonlinear polaritonic dynamics in hBN. Under these conditions, the two atoms in each unit cell (B and N) move in the same way across the crystal, and therefore, we only need to consider a central unit cell with the atom label $l$ taking the values B or N. Clearly, the total potential (the quantity in square brackets in the right-hand side of eq\ \ref{newton1}) only depends on the relative coordinate $\ub=\ub_{\rm N}-\ub_{\rm B}$, whereas the center of mass moves at constant velocity. We thus have $\ub_{\rm B}=-(M/M_{\rm B})\ub\approx-0.564\,\ub$ and $\ub_{\rm N}=(M/M_{\rm N})\ub\approx0.436\,\ub$, where $M=M_{\rm B}M_{\rm N}/(M_{\rm B}+M_{\rm N})\approx6.102\,$Da is the reduced mass (assuming naturally abundant isotope distributions), while the equation of motion becomes
\begin{align}
M\left[\ddot{\ub}(t)+\tau^{-1}\dot{\ub}(t)\right]=-\nabla_{\ub}\left[\bar{\mathcal{E}}(\ub)+\int_{\rm UC} d^3\rb\;\rho(\ub,\rb)\;\phi^{\rm ext}(\rb,t)\right].
\label{newton2}
\end{align}
Here, $\rho(\ub,\rb)$ has the periodicity of the crystal, so the integral only extends over one unit cell (UC). Also, $\bar{\mathcal{E}}(\ub)$ is the internal energy per unit cell, which we obtain from DFT (see below). Finally, based on the smallness of the wave vectors accessed through external illumination, we can approximate $\phi^{\rm ext}(\rb,t)=-\Eb^{\rm ext}(t)\cdot\rb$ in terms of an external electric field $\Eb^{\rm ext}(t)$, which allows us to rewrite eq\ \ref{newton2} as in eq\ \ref{newton3}, where $\bar{\pb}(\ub)=\int_{\rm UC} d^3\rb\;\rho(\ub,\rb)\;\rb$ is the unit cell dipole. In this work, we simulate the nonlinear response of monolayer hBN in the in-plane phonon spectral region by solving eq\ \ref{newton3} together with the DFT-based parametrization given in eqs\ \ref{Epu} for $\bar{\mathcal{E}}(\ub)$ and $\bar{\pb}(\ub)$.

%10B: M=10.0129369 amu, 19.9%  of atoms
%11B: M=11.0093052 amu, 80.1%  of atoms -> average B: M=10.8110 amu
%14N: M=14.0030740 amu, 99.64% of atoms
%15N: M=15.0001089 amu,  0.36% of atoms -> average N: M=14.0067 amu   -> M(with averages) = 6.10155 amu

% --- DFT -------------------------------------------------
\section{DFT Calculations} We use the Vienna \textit{ab initio} simulation package (VASP) \cite{KF96,KH93,KF96_2} to carry out first-principles DFT calculations using the projector-augmented-wave (PAW) method \cite{B94_2} together with the generalized gradient approximation of Perdew-Burke-Ernzerhof (GGA-PBE) for the exchange-correlation functional \cite{PBE96}. A vacuum spacing of 10 \AA{} between adjacent images is introduced to prevent artificial interactions. The plane-wave energy cut-off is set to 500 eV. We use the conjugate gradient method to optimize the structure with an energy convergence criterion of $10^{-8}\,$eV between two ionic steps. A $\Gamma$-centered wave-vector sampling grid of size $18\times18\times1$ is used for the structural relaxation. Atomic positions and lattice vectors are relaxed until the total force in the unit cell is reduced to a value below $10^{-7}\,$eV/{\AA}. The calculated B-N bond distance differs by just $+0.3\%$ from the measured value in bulk hBN. To obtain the charge density $\rho(\ub,\rb)$, we calculate the electron densities corresponding to each atomic displacement using a sufficiently dense grid in the unit cell. The contribution from the nuclei and the K-shell electrons is incorporated by assimilating them to point charges ($3e$ for B and $5e$ for N), which add a term $\bar\pb^{\rm nucl}(\ub)=13ea\,\xx+eM(5/M_{\rm N}-3/M_{\rm B})\,\ub$ to the unit cell dipole $\bar\pb(\ub)=\bar\pb^{\rm el}(\ub)+\bar\pb^{\rm nucl}(\ub)$, where $\bar\pb^{\rm el}(\ub)$ originates in the 8 outer electrons per unit cell. The result from this analysis for in-plane atomic displacements is well described by the polynomial expressions in eqs\ \ref{Epu}, which are compatible with the crystal symmetries of the hBN monolayer. The linear phonon frequencies and the dynamical matrix are calculated using the small displacement method.

% --- Table peaks -----------------------------------------
\begin{table*}[t]
\caption{{\bf Peak nonlinear susceptibility in the perturbative regime.} We focus on polarization along the $x$ direction (parallel to B-N bonds) and consider different incoming light frequencies $\omega_{\rm in}$, harmonics $s\omega_{\rm in}$, and perturbation orders $n$. The $\Delta\omega_{\rm in}$ column gives the FWHM of the $\big|\chi^{ns}_{\omega_{\rm in}}\big|^2$ as a function of incident frequency. The rightmost column shows the corresponding values of the susceptibilities in SI units for $\tau=2\,$ps.}
\begin{tabular}{l c c | c | c | c}
\\
$\omega_{\rm in}$ &  $(n,s)$ & $s\omega_{\rm in}$ & $\chi^{ns}_{\omega_{\rm in}}$  & $\Delta\omega_{\rm in}$ &  $|\chi^{ns}_{\omega_{\rm in}}|$\,(m$^{n-1}$/V$^{n-1}$) \\
\hline
& & & & & \\
$\omega_0$ & (1,1) & $\omega_0$ & $\propto\tau$ & $1/\tau$ & $14.8$ \\
& & & & & \\
$\omega_0$ & (2,2) & $2\omega_0$ & $\propto\tau^2$ & $\approx0.64/\tau$ & $6.88\times10^{-10}$ \\
& & & & & \\
$\omega_0$ & (3,3) & $3\omega_0$ & $\propto\tau^3$ & $\approx1/2\tau$ & $2.97\times10^{-18}$ \\
& & & & & \\
$\omega_0$ & (3,1) & $\omega_0$ & $\propto\tau^4$ & $\approx0.44/\tau$ & $3.22\times10^{-15}$ \\
& & & & & \\
$\omega_0/2$ & (2,2) & $\omega_0$ & $\propto\tau$ & $1/2\tau$ & $4.37\times10^{-11}$ \\
& & & & & \\
$\omega_0/2$ & (3,3) & $3\omega_0/2$ & $\propto\tau$ & $1/2\tau$ & $9.73\times10^{-23}$ \\
& & & & & \\
$\omega_0/3$ & (3,3) & $\omega_0$ & $\propto\tau$ & $1/3\tau$ & $1.30\times10^{-22}$ \\
& & & & & \\
\end{tabular}
\label{Table1}
\end{table*}

% --- perturbative ----------------------------------------
\section{Perturbation Limit} We find a perturbative solution to eq\ \ref{newton3} by expanding the displacement vector as $\ub=\sum_{ns}\ub^{ns}\ee^{-\ii s\omega t}$, where $n$ is the scattering order and $s$ is the harmonic, subject to the condition $|s|\le n$. For a monochromatic field $\Eb^{\rm ext}(t)=2\Eb_0\cos(\omega t)$, this leads to the recurrence relation
\begin{align}
M\,\big[\omega_0^2-s\omega(s\omega+\ii\tau^{-1})\big]&\,u_i^{ns}
=-e_1\sum_{jk}\sum_{n's'}a_{ijk}u_j^{n's'}u_k^{n-n',s-s'} \nonumber\\
&-e_2\sum_{jkl}\sum_{n's'n''s''}b_{ijkl}u_j^{n's'}u_k^{n''s''}u_l^{n-n'-n'',s-s'-s''} \nonumber\\
&+Q_0\delta_{n,1}\big(\delta_{s,1}E_{0,i}+\delta_{s,-1}E_{0,i}^*\big) \nonumber\\
&+Q_1\sum_{j}c_{ijk}\big(u_j^{n-1,s-1}E_{0,k}+u_j^{n-1,s+1}E_{0,k}^*\big) \nonumber\\
&+Q_2\sum_{jkl}\sum_{n's'}d_{ijkl}u_j^{n's'}\big(u_k^{n-n'-1,s-s'-1}E_{0,l}+u_k^{n-n'-1,s-s'+1}E_{0,l}^*\big), \nonumber
\end{align}
where $i,j,k,l\in\{x,y,z\}$ denote Cartesian components and the only nonzero coefficients inside the sums (extracted from eqs\ \ref{Epu}) are $a_{xxx}=-a_{yxy}/2=-a_{xyy}=3$, $b_{xxxx}=b_{xxyy}=b_{yyxx}=b_{yyyy}=4$, $c_{xxx}=-c_{yyx}=-c_{xyy}=-c_{yxy}=2$, $d_{xxxx}=d_{yyyy}=3$, $d_{yxyx}=d_{xxyy}=2$, and $d_{xyyx}=d_{yxxy}=1$. Incidentally, this equation leads to the vanishing of $u_i^{ns}$ if $n+s$ is odd. Through iterative solution, we obtain the displacement components $u_i^{ns}$, which upon insertion into eq\ \ref{pu}, also produce analytical expressions for the induced dipole. In particular, for incident polarization along $x$, the displacement vectors are found to be confined along $x$ as well, and we obtain the perturbation series
\begin{align}
\frac{p_x(t)}{\epsilon_0\mathcal{V}}=\sum_{s}\chi^{(s)}_\omega E_0^s \ee^{-\ii s\omega t} \nonumber
\end{align}
for the polarization density, where we define the field-dependent harmonic susceptibilities $\chi^{(s)}_\omega=\sum_n \chi^{ns}_\omega |E_0|^{n-s}$, which should coincide with eq\ \ref{chisw}. For incidence frequency near $\omega_0$, the leading terms in the perturbation series are those involving the higher powers in the resonant factor $\big[\omega_0^2-\omega(\omega+\ii\tau^{-1})\big]^{-1}$, namely,
\begin{subequations}
\begin{align}
\chi^{11}_{\omega}&\approx\frac{Q_0^2}{\epsilon_0\mathcal{V}M}\frac{1}{\omega_0^2-\omega(\omega+\ii\tau^{-1})}, \\
\chi^{22}_{\omega}&\approx\frac{Q_0^3}{\epsilon_0\mathcal{V}M^2}\bigg(\frac{e_1}{M\omega_0^2}+\frac{Q_1}{Q_0}
\bigg)\frac{1}{\big[\omega_0^2-\omega(\omega+\ii\tau^{-1})\big]^2}, \\
\chi^{33}_{\omega}&\approx\frac{\ii Q_0^4}{4\epsilon_0\mathcal{V}M^4\omega_0^2}\bigg(
2e_2+\frac{8e_1Q_1}{Q_0}+\frac{3e_1^2}{M\omega_0^2}+\frac{4MQ_2\omega_0^2}{Q_0}
\bigg)\frac{1}{\big[\omega_0^2-\omega(\omega+\ii\tau^{-1})\big]^3}, \\
\chi^{31}_{\omega}&\approx\frac{6Q_0^4}{\epsilon_0\mathcal{V}M^4}\bigg(\frac{5e_1^2}{M\omega_0^2}-2e_2\bigg)\frac{1}{\big[\omega_0^2-\omega(\omega+\ii\tau^{-1})\big]^2\big|\omega_0^2-\omega(\omega+\ii\tau^{-1})\big|^2}. \label{chi31}
\end{align}
\end{subequations}
These expressions are in excellent agreement with the numerical solution of the equation of motion (eq\ \ref{newton3}) for low field intensity (see Figure\ \ref{Fig2}). Their scaling with the lifetime $\tau$ is summarized in Table\ \ref{Table1}, along with explicit values at $\omega=\omega_0$.

Incidentally, for the on-resonance Kerr effect, we have $\chi^{(1)}_{\omega_0}=\chi^{11}_{\omega_0}+\chi^{31}_{\omega_0}|E_0|^2+\chi^{51}_{\omega_0}|E_0|^4+\dots$, contributed by odd-order susceptibilities
\begin{align}
\chi^{11}_{\omega_0}&=\frac{iQ_0^2\tau}{\epsilon_0\mathcal{V}M\omega_0}, \nonumber\\
\chi^{31}_{\omega_0}&=-\frac{6Q_0^4\tau^4}{\epsilon_0\mathcal{V}M^4\omega_0^4}\left(5e_1^2/M\omega_0^2-2e_2\right), \nonumber\\
\chi^{51}_{\omega_0}&=-\frac{36iQ_0^6\tau^7}{\epsilon_0\mathcal{V}M^7\omega_0^7}\left(5e_1^2/M\omega_0^2-2e_2\right)^2. \nonumber\\
\end{align}
The $n=3$ term is $90^\circ$ out of phase with respect to $n=1$, and thus, these two do not interfere in the resulting intensity $I^{(1)}\propto|\chi^{(1)}_{\omega_0}|^2$. Actually, for low $E_0$, the $n=3$ contribution produces an increase in $I^{(1)}$, which is however compensated by the $n=5$ term (see Figure\ \ref{Fig4}a and discussion in the main text). Nevertheless, for sufficiently off-resonance $\omega$, the perturbative result embodied in eq\ \ref{chi31} reproduces the decay of the curves in Fig.\ \ref{Fig2}(b) away from the resonance (not shown).

% --- phonon quantization ---------------------------------
\section{Phonon Quantization and Quantum Blockade} In the linear regime, the potential energy can be approximated by the quadratic expression $\mathcal{E}(\{\ub\})=(1/2)\sum_{ll'}\ub_l\cdot\mathcal{D}_{ll'}\cdot\ub_{l'}$, so in the absence of external illumination, the solution to the classical equation of motion (eq\ \ref{newton1}) admits an expansion $\ub_l=M_l^{-1/2}\sum_n c_n\eb_{nl}$ in terms of a complete ($\sum_n\eb_{nl}\otimes\eb_{nl'}=\delta_{ll'}\mathcal{I}_3$) and orthonormal ($\sum_l\eb_{nl}\cdot\eb_{n'l}=\delta_{nn'}$) basis set of eigenvectors $\eb_{nl}$ of the real, symmetric dynamical matrix (i.e., $\sum_{l'}(M_lM_{l'})^{-1/2}\mathcal{D}_{ll'}\cdot\eb_{nl'}=\omega_n^2\eb_{nl}$, where $\omega_n$ are real oscillation eigenfrequencies). In a quantum description, we can write the Hamiltonian associated with atomic vibrations in general as $\hat{\mathcal{H}}=-\sum_l\hbar^2\nabla_{\ub_l}^2/2M_l+\mathcal{E}(\{\ub\})=(1/2)\sum_n\big(-\hbar^2\nabla^2_{c_nc_n}+\omega_n^2c_n^2\big)$, where the rightmost expression, obtained by replacing the displacement vectors by the expansion coefficients $c_n=\sum_l\sqrt{M_l}\eb_{nl}\cdot\ub_l$, consists of a sum over quantum harmonic oscillators. Following a standard second quantization procedure, we interpret $c_n$ and $-\ii\hbar\partial_{c_n}$ as displacement and momentum operators, respectively, from which we define phonon creation and annihilation operators $\hat{b}_n^\dagger$ and $\hat{b}_n$ through $c_n=\sqrt{\hbar/(2\omega_n)}(\hat{b}_n^\dagger+\hat{b}_n)$ and $-\ii\hbar\partial_{c_n}=\ii\sqrt{\hbar\omega_n/2}(\hat{b}_n^\dagger-\hat{b}_n)$, in terms of which the Hamiltonian reduces to $\hat{\mathcal{H}}=\hbar\sum_n\omega_n(\hat{b}_n^\dagger\hat{b}_n+1/2)$. The rms displacement of atom $l$ associated with the presence of one phonon in mode $n$ is thus given by $\sqrt{\langle0|\hat{b}_n|\ub_l|^2\hat{b}_n^\dagger|0\rangle}=\sqrt{3\hbar/(2M_l\omega_n)}\,|\eb_{nl}|$, where we have used the noted expansion of $\ub_l$ in terms of eigenmodes, and the expansion coefficients $c_n$ in terms of ladder operators. For a hBN flake consisting of a finite number $N$ of unit cells, approximating the eigenvectors to those of an infinite crystal, orthornormality implies $|\eb_{nl}|\sim1/\sqrt{N}$, and in particular, for oscillations at the $\Gamma$ point, the rms displacement associated with one quantum reduces to $\sqrt{\langle u^2\rangle}\approx\sqrt{3\hbar/(2NM\omega_0)}$.

% --- nonlinear oscillation frequency ---------------------
\section{Oscillation Frequency beyond the Linear Regime} Focusing on atomic motion along $x$, we find the self-sustained oscillation frequency by multiplying eq\ \ref{newtonx} by the velocity $\dot{u}_x$, neglecting losses ($\tau^{-1}=0$) and setting the external drive to zero ($E_0=0$). As a function of the maximum displacement $u_x^{\rm max}>0$ toward the positive $x$ direction, where the potential is lower (see Figure\ \ref{Fig1}a), direct integration then yields the oscillation period $T=2\int_{u_x^{\rm min}}^{u_x^{\rm max}}du_x/\sqrt{f(u_x^{\rm max})-f(u_x)}$, where $f(u_x)=\omega_0^2u_x^2+(2e_1/M)\,u_x^3+(2e_2/M)\,u_x^4$ and $u_x^{\rm min}<0$ is the lower bound of the oscillation defined by $f(u_x^{\rm min})=f(u_x^{\rm max})$. We find $u_x^{\rm min}$ from the analytical solution of resulting polynomial of the fourth degree, and then numerical integrate the above expression to obtain the oscillation frequency $\omega=2\pi/T$, plotted in Figure\ \ref{Fig4}b as function of $u_x^{\rm max}$.

% Direct application of the methods used above allows us to derive the relations $\bar{\alpha}^{(2)}(\omega)\equiv\bar{\alpha}^{(2)}_{xxx}(\omega)=(-A/C)\bar{\alpha}^{(2)}_{xyy}(\omega)=(-A/2C)\bar{\alpha}^{(2)}_{xxy}(\omega)$ for the nonzero polarizability components (defined such that $\alpha_{ii'i''}E_{0,j}E_{0,i'}$ gives the induced dipole along a direction $i=x,y$), the maximum second-order response is produced for right- and left-circularly polarized light, giving rise to left- and right-circularly polarized second-harmonic emission, respectively, with an intensity that exceeds by a factor of 2 the one obtained for the same incident intensity and linear polarization along $x$. In contrast, with $y$ polarization, SHG is forbidden due to the lattice symmetry.

% --- reflected power -------------------------------------
%{\bf Reflected Power.} The reflected power is constructed from the emission of a periodic array of dipoles of strength $\bar{\alpha}^{(s)}(\omega)(E_0)^s$ per unit cell. The corresponding lattice sum leads to the reflected power \cite{paper090}
%\begin{align} I^{(s)}_{\rm ref}=\frac{c}{2\pi}\bigg|\frac{2\pi s\omega}{\mathcal{A}c} \bar{\alpha}^{(s)}(\omega)(E_0)^s\bigg|^{2} \nonumber \end{align}
%at each harmonic frequency $s\omega$ for incident frequency $\omega$ and field amplitude $E_0$. Here, $\mathcal{A}=(3\sqrt{3}/2)\,a^2=5.43\,{\AA}^2$ is the unit cell area.

\end{widetext} %---arxiv

% =========================================================
% --- SI, acknowledgments, bibliography, etc. -------------
% =========================================================

% --- SI --------------------------------------------------
%\section*{SUPPLEMENTARY MATERIALS} %---Sci Adv optional
%\section*{Supplementary Information} %---ACS optional
%\noindent Supplementary material for this article is available at http://advances.sciencemag.org/cgi/content/full/...
%The Supporting Information is available free of charge at https://pubs.acs.org/doi/xxx. %---ACS optional

% --- acknowledgments -------------------------------------
%\section*{Funding} % ... text ... %---OSA, place funding here and thanks to colleagues below
\section*{ACKNOWLEDGEMENTS} %---ACS---arxiv
%\section*{Acknowledgments} %---OSA
%\begin{acknowledgments} %---APS
%\noindent {\bf Funding:}
This work has been supported in part by the European Research Council (Advanced Grant 789104-eNANO), the Spanish MINECO (Severo Ochoa CEX2019-000910-S), the Catalan CERCA Program, and Fundaci\'{o}s Cellex and Mir-Puig. %---ACS
%{\bf Competing interests:} The authors declare that they have no competing interests. {\bf Data and materials availability:} All data needed to evaluate the conclusions in the paper are present in the paper and/or the Supplementary Materials. Additional data related to this paper may be requested from the authors. %---ACS---OSA---arxiv
%\end{acknowledgments} %---APS
%\section*{Disclosure} The authors declare no conflicts of interest. %---OSA

% --- bibliography (adapt file path as appropriate) -------
%\bibliographystyle{apsrev} %---APS---SI---arxiv %---comment out for longbibliography
%\bibliographystyle{ScienceAdvances} %--- Science
%\bibliography{../../../bibtex/refs.bib} %---APS---OSA---SI---Science---arxiv with lower-case title format
%\bibliography{../../../bibtex/refsU.bib} %---ACS with upper-case title format

%\clearpage %--- optional
%\pagebreak \onecolumngrid \section*{SUPPLEMENTARY FIGURES} %---SI---arxiv optional

\end{document}